\documentclass{aa}  
\usepackage{comment}
\usepackage{xcolor}
\usepackage{soul}
\usepackage{subfig}

\usepackage{ulem}
\usepackage{tikz}

\def\gsim{\lower.5ex\hbox{$\; \buildrel > \over \sim \;$}}
\def\lsim{\lower.5ex\hbox{$\; \buildrel < \over \sim \;$}}

\usepackage{ulem}

\usepackage{graphicx}
\usepackage{txfonts}
\usepackage{hyperref}

\begin{document} 

   \title{Very High Energy Gamma Rays from Ultra Fast Outflows}

   \author{B. Le Nagat Neher
          \inst{1}
          \and
          E. Peretti\inst{2,3}\fnmsep
          \and
          P. Cristofari  \inst{1}
          \and 
          A. Zech \inst{1}
          }

   \institute{LUX, Observatoire de Paris, Université PSL, Sorbonne Université, CNRS, 92190 Meudon, France
         \and
    INAF, Istituto Nazionale di Astrofisica, Osservatorio Astronomico di Arcetri, Largo E. Fermi 5, 50125 Florence, Italy
        \and
    APC, Universit\'e Paris Diderot, CNRS/IN2P3, CEA/Irfu, Observatoire de Paris, Sorbonne Paris Cit\'e, France
    }

   \date{Received XXXX XX, ; accepted March 8, 2026}

 
  \abstract
  {Ultra fast outflows (UFOs) from active galactic nuclei (AGN) are expected to lead to the formation of  sub-relativistic strong shocks expanding in a  dense circumnuclear medium, and thus  have the potential for being efficient particle accelerators, and to be proficient sources of gamma rays and neutrinos. }
   {We investigate the detectability of a sample of nearby identified UFOs in gamma rays and neutrinos with current and next-generation instruments.}
   {We model the acceleration of particles at the strong shocks of UFOs, and estimate the associated gamma-ray and neutrino signal. We adopt our model to investigate the prospects for detection with current and next-generation observatories.}
   {We find that several UFOs could be detectable in the very-high-energy (VHE) domain - for example, by the Cherenkov Telescope Array Observatory (CTAO)- even if they remain undetected by Fermi-LAT in the high-energy range. Detectability is favored for hard proton spectra (spectral index $\alpha \lesssim$ 3.9), high acceleration efficiencies, and amplified magnetic fields. Our results suggest that next-generation VHE observatories could detect the first gamma-ray signatures of AGN UFOs, providing a new probe of particle acceleration in sub-relativistic shocks.}
   {}


   \maketitle

\section{Introduction}

Active galactic nuclei (AGN) are powered by accretion onto supermassive black holes (SMBHs) and they are considered to play a role in feedback on galaxies~\citep{Feedback_Silk_1998}.
Thanks to advances in X-ray spectroscopy, fast ionized outflows were detected more than 20 years ago~\citep{Pounds_outflows_2003,Reeves_r_ray_outflow_2003}.
These outflows have been identified after observations of blueshifted Fe absorption lines~\citep{Chartas_2002,relativistic_outflow_2009}.

The typical distances on which these outflows expand span from fractions of a parsec to tens of parsecs from the SMBH~\citep{SMB_winds_x_rays_2024}, with typical velocities
 $v\geq 0.1\rm c$ and a wide range of mass-loss rate, $\dot{M}\approx 10^{-4}-100 \, \rm M_{\odot} \, yr^{-1}$.
Wind velocities up to $v\approx0.4\rm c$ have been observed by~\citet{gofford_statistics_suzaku} and even a record measurement at $v=0.76\rm c$ has been reported~\citep{speed_record_Chartas_2009}.
Such high velocities granted these winds the current name of Ultra Fast Outflows~\citep[UFO,][]{UFO_evidence_tombesi_2010,UFO_observations_feedback_2015}. 
UFOs are indeed sub-relativistic winds launched with a wide opening angle, within which shocks can form, typically as a consequence of their fast expansion in the interstellar medium (ISM).
Regardless of the AGN class, Radio-loud~\citep[jetted,][]{2014UFO_AGNRL} or Radio-quiet~\citep{2012UFO_AGNRQ}, UFO detections with common properties~\citep{Radio_L_and_Q_AGN_winds_Mestici_2024} have been observed. This suggests that UFOs may be relatively common in AGNs, independent of the presence of a co-existing jet.
The launching mechanisms of UFOs remain unclear. 
They could be radiation-driven~\citep{radiation_mhd_agn_disk_2014} or magnetically-driven outflows~\cite{UFO_launching_Fukumura_2010,UFO_launching_Fukumura_2014,SMB_winds_x_rays_2024}. 
The geometry and the environments of AGN-driven winds have been explored by~\cite{2012fauchergiguere} and~\cite{UFO_signature_Nims} among others. 
However, the details of the circumnuclear medium properties in which UFOs expand are still poorly known.

The physical conditions met at UFOs, in particular their large speed and kinetic power, make them particularly suited for diffusive shock acceleration (DSA), with the formation of strong sub-relativistic collisionless shocks in a dense medium. 
The energization of particles at least up to the very-high-energy (VHE) domain 
is expected, in turn, to lead to the production of gamma rays and neutrinos in the high-energy (>1 GeV) and VHE range (>1 TeV).  
Observational support for particle acceleration in UFOs has come from a stacking analysis with Fermi-LAT of 11 selected UFO hosts~\citet{gamma_UFO_Ajello_2021}. This work showed that UFOs are a class of gamma-ray emitters.

\begin{figure}
   \centering
   \includegraphics[scale=0.70]{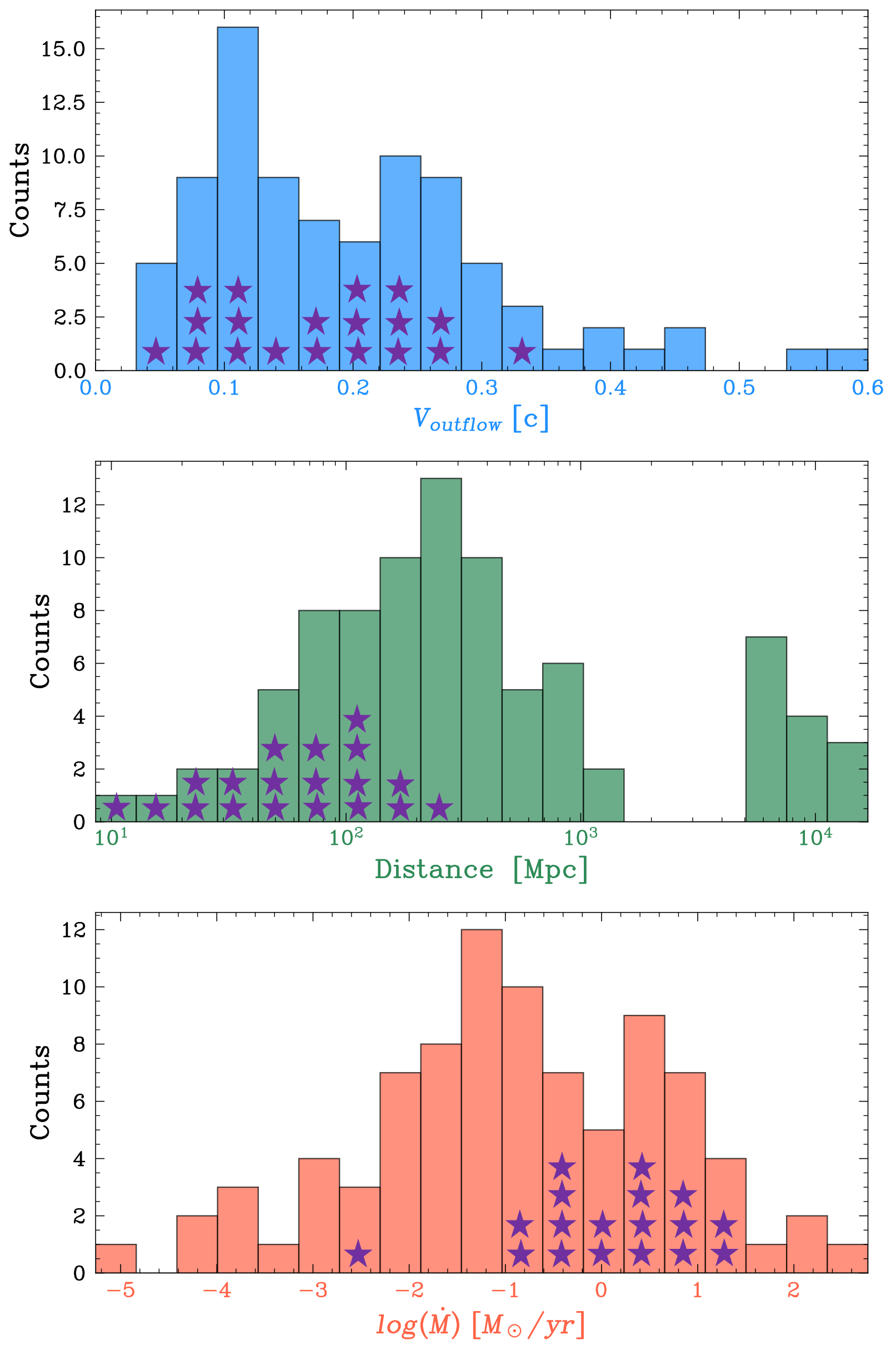}
      \caption{Description of the sample of sources, top panel : distributin of the outflow speed, on the middle the distance and on the bottom one the mass-loss rate, in violet stars are indicated the main properties of the detectable UFOs, summarized in Table \ref{tab:names_fig2}.}
         \label{source sample}
   \end{figure}

A definitive detection of UFOs in the gamma-ray domain remains elusive. The nearby UFO host NGC~4151 is spatially consistent with a Fermi-LAT source~\citep[]{Murase_Seyfert,ngc4151}, hinting at possible gamma-ray emission from the UFO, though the association is not yet beyond doubt due to a possible contamination from a blazar at 5 arcmin.
NGC~1068, despite its high column density, is also a plausible UFO host detected in the GeV range. However, the origin of such gamma rays is highly uncertain due to the co-existence of many possible acceleration sites such as star-forming regions \citep{Yoast-Hull_2014,Eichmann2022,Ajello2023}, an AGN molecular outflow \citep{Lamastra2016,Lamastra2019}, a mildly relativistic radio jet~\citep{Lenain2010,Slavatore2024} ~\citep[see also][for a detailed review]{Padovani2024}.
More than 15 years of Fermi-LAT monitoring have otherwise yielded upper limits to individual UFO hosts.
UFOs have also been proposed and studied as potential sources of ultra-high-energy cosmic rays observed at Earth~\citep{peretti2023diffusive,Ehlert_2024_UFO_UHECR}.
Here we study the gamma-ray and neutrino emissions for a list of selected UFOs, reported after the detection of blueshifted absorption lines of highly ionized iron in their X-ray spectrum obtained with XMM-Newton~\citep{Ehlert_2024_UFO_UHECR} whose physical properties have been summarized in Fig.\ref{source sample}.

We discuss the prospects for detection with current and next-generation observatories, for gamma rays and for neutrinos.  
In this work, we focus on an order-of-magnitude discussion, that encompasses all shocks developed throughout the UFO expansion, on the prospects for detection in gamma rays in the GeV, TeV and 100 TeV range, accounted by the typical sensitivities of Fermi-LAT, the CTAO and LHAASO~\citep{Atwood_2009_Fermi_paper,2018_CTAO_book,Neronov_2020_LHAASO_paper}.
We focus on the hadronic component of the gamma-ray emission. The leptonic contribution is expected to be sub--dominant in the conditions relevant for UFOs. On the one hand, the high ambient densities ($\sim 10^2-10^5~\mathrm{cm} ^{-3}$~\citep{laha2019physicsUFO,2023density_ricci}) strongly enhance the efficiency of hadronic interactions and subsequent $\pi^0$-decay gamma-ray production. On the other hand, the magnetic fields inferred for these environments lead to severe synchrotron losses for relativistic electrons, limiting both their maximum attainable energy and their radiative efficiency in the gamma-ray domain. Moreover, a leptonic gamma-ray component strong enough to dominate would inevitably imply bright synchrotron counterparts from radio- to X-ray wavelengths, which are generally not observed in association with such outflows. As a result, the hadronic channel is expected to provide the dominant contribution to the high-energy emission.

Regardless of the cases of NGC~4151 and NGC~1068, UFOs have not been clearly detected by Fermi-LAT \citep{Fermifourthcatalog}, we are looking to investigate UFO contributions in the TeV range that could be detectable by the CTAO, offering typically a tenfold improvement in TeV range sensitivity over current atmospheric Cherenkov telescopes. 
The prospects for detection in neutrinos is discussed considering the sensitivity of KM3NeT/ARCA~\citep{KM3NeT} noting that an analogous comparison could be done with IceCube-Gen2~\citep{Ice_Cube_gen2}.

The manuscript is organized as follows, in section \S~\ref{Section: Model} we introduce the framework of our investigation describing the main scalings of UFOs and the computation of the associated gamma-ray and neutrino fluxes. In Section \S~\ref{Section: Results} we present our results on the UFO detectability with gamma rays and neutrinos. 
We discuss our findings and draw our conclusions in \S~\ref{Section: Conclusions}.

\section{Gamma rays and neutrinos from UFOs}
\label{Section: Model}
\subsection{Particle acceleration at UFOs}
The launching of UFOs driven by AGN is expected to lead to the formation of strong shocks (Mach number $\mathcal {M} >>1$). 
In the initial phase, the outflow freely expands into the surrounding circumnuclear medium leading to the formation of a forward shock (FS). 
The deceleration of the wind material due to the collision with the external medium gives rise to the formation of a second shock, the wind termination shock (TS), inside the wind structure. These two shocks are coupled in the first stages of the dynamics of the wind. 
When the swept-up external medium material balances the wind mass itself, the expansion slows down and the pressure equilibrium leads to the decoupling of the two shocks~\citep{weaver1977,chevalier1985}. The deceleration time $t_{\rm dec}$ at which this happens reads:
   \begin{equation}
    t=t_{\rm dec} \implies \dot{M}t=\frac{4\pi}{3}R_{\rm FS}^3(t)\rho_0,
   \end{equation}
where $\dot{M}$ is the mass outflow rate, $R_{\rm FS}$ is the forward shock radius and $\rho_0$ is the density of the surrounding medium.
We stress that our model does not describe the propagation of a single observed UFO clump from sub-parsec scales out to distances of 1-10 pc. 
Rather, we assume that the kinetic power injected by the outflow drives strong collisionless shocks in the surrounding circumnuclear medium. 
The characteristic size of the shock region is treated as a free parameter, reflecting the uncertain geometry and dynamical evolution of the wind–ISM interaction. Importantly, the duration of an observed UFO episode does not determine the lifetime of the shock.
Even a short-lived outflow as short as $t_{\rm obs}\sim 10^5\,\rm s$ in PDS456 \citep{2025XU_PDS456} can deposit sufficient kinetic energy to launch a shock that propagates over much longer timescales $t_{\rm wind}\sim 10^4\,\rm yr$ and distances, as commonly observed in transient astrophysical systems (e.g., supernovae or novae). While the transient nature of UFO observations introduces uncertainty on the the wind properties, the small dispersion of UFO characteristics reported in the literature across multiple observations of individual sources \citep[e.g.,][]{UFO_evidence_tombesi_2010,2023Matzeu,2021Chartas} indicates that even a single X-ray detection can provide meaningful insight into their potential as VHE gamma-ray emitters. Our treatment therefore models the long-term interaction between the integrated wind kinetic power and the ambient medium, rather than the direct propagation of an individual UFO clump to parsec scales.

The deceleration time can thus be expressed as follows:
   \begin{equation}
    t_{\rm dec} \approx 56 \left(  \frac{\dot{M} }{1 \rm M_\odot/yr}\right)^{{1}/{2}} \left(\frac{v_{\infty}}{0.1~\rm c}\right)^{-3/2} \left(\frac{n_0}{100 \, \rm ~cm^{-3}} \right)^{-{1}/{2}}~ ~  ~ ~\text{yr},
   \end{equation}
where $n_0 = \rho_0/m_{\rm p}$ and $v_{\infty}$ is the terminal wind speed.

If the source age satisfies $t_{\rm age} \lesssim t_{\rm dec}$ the FS and TS remain coupled and the former can be expected to be the only strong shock of the system. 
For $t_{\rm age}\gtrsim t_{\rm dec}$, they can be considered to be distinct and, at late time ($t_{\rm age}\gg t_{\rm dec}$), the FS is expected to be slow and radiative, thus limiting particle acceleration up to the highest energies.

At these shocks, depending on the stage of the evolution, strong collisionless shock conditions can be found. Thus DSA is expected to be efficient in energizing particles. 
Indeed, the large wind velocities, typically $\gtrsim 0.1 $c, ensure that in all dynamical phases, at least one shock's Mach number is $\mathcal{M} \gg 1$, while the shocks remains sub-relativistic, thereby allowing for particles to get injected efficiently in the acceleration process. 

A priori, particle acceleration at the TS and FS can differ, as the different acceleration conditions concur in setting the maximum momentum and the normalization of the accelerated particles.

\begin{figure}
   \centering
   \includegraphics[scale=0.35]{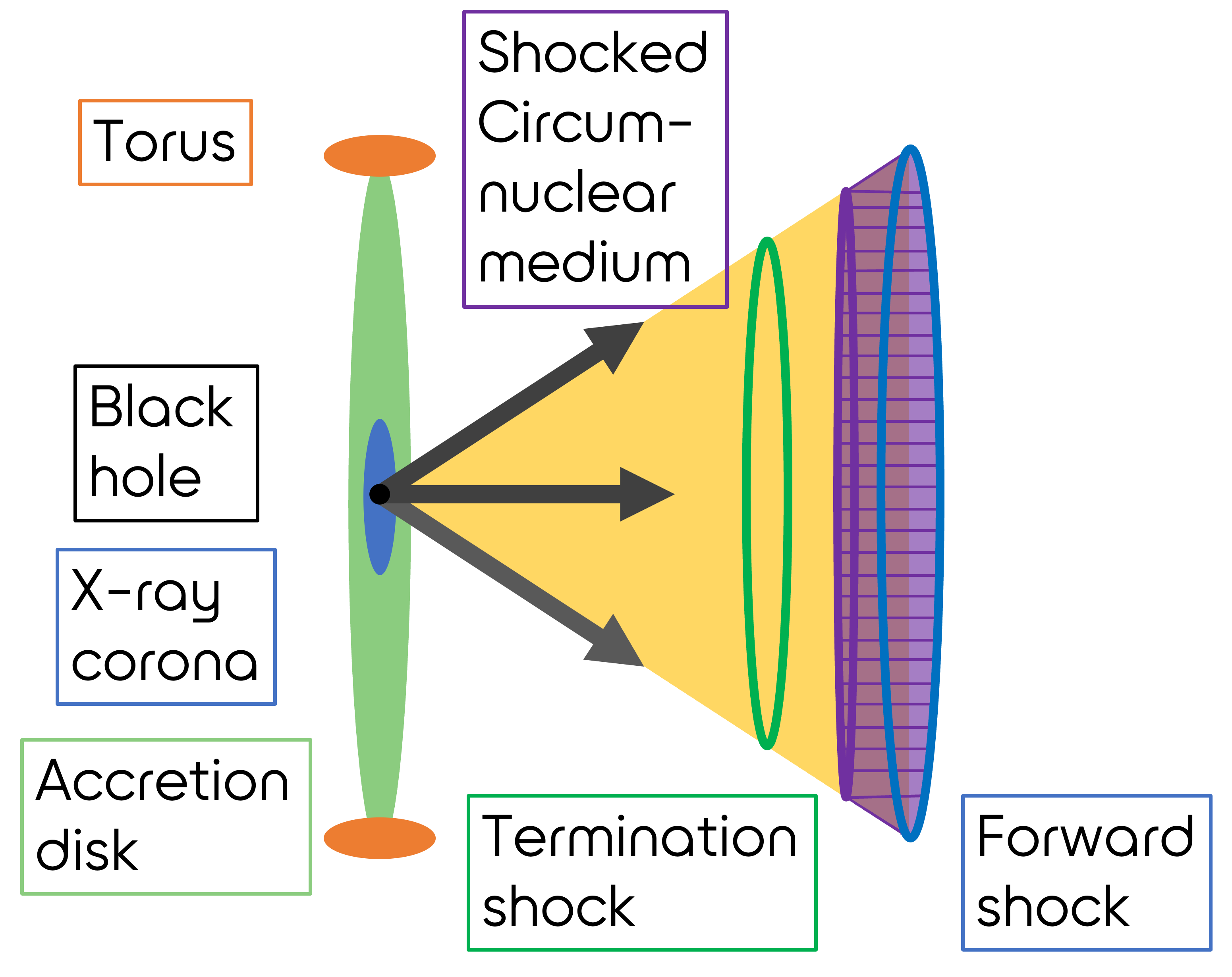}
      \caption{Schematic representation of the geometry of UFOs and associated shocks.}
         \label{fig:schematic_view}
   \end{figure}

Our modeling implicitly assumes that shocks form as the UFO interacts with the surrounding circumnuclear medium. While direct observational evidence for such shocks is still lacking, physical arguments make their formation difficult to avoid. Indeed for an outflow with characteristic velocity $v_{\infty}\sim0.1\rm c$, shock formation is expected whenever the relative velocity exceeds the magnetosonic speed of the ambient medium, $v_{\rm ms}=(c_s^2+v_A^2)^{1/2}$. Avoiding shocks would therefore require unusually extreme external conditions, such as temperatures $T\sim10^{10}\,\rm K$ or magnetic field strengths close to $B\sim 0.1\rm  \,G$, larger than those typically inferred at circumnuclear scales \citep{2004faraday_rotation}. In the absence of such conditions, the interaction between the UFO and its environment is expected to generically produce strong shocks.
A schematic view of the geometry under consideration is shown in Fig.~\ref{fig:schematic_view}. To account for potential acceleration at both the FS and the TS, we normalize our spectrum using assumptions that can be easily translated from one case to the other.

As mentioned above, depending on the evolutionary stage of the UFO, one expects the FS or the TS to be the most relevant site for DSA. Therefore, we work under the assumption that DSA is at play at one shock, without distinguishing between one or the other.
We consider that protons are accelerated through DSA, and that their distribution follows a power-law in momentum:
\begin{equation}
f(p)=
A
\left(\tfrac{p}{p_0}\right)^{-\alpha}
\exp\!\left(-\tfrac{p}{p_{\rm max}}\right)
\label{protons distrib}
\end{equation}
where $\alpha$ the index of the spectrum and $p_{\rm max}$ the maximum momentum deduced from the Hillas criterion (see discussion below).
In the test-particle limit of DSA, at strong shocks, $\alpha =4$, but several non-linear effects can lead to deviations from the test-particle case and produce harder/steeper spectra~\citep{malkov2001,amato2005,zirakashvili2008,caprioli2020,cristofari2022}. 
The normalization $A$ has been obtained  assuming that a fraction $\xi_{\rm CR}$ of the fluid ram pressure at the shock,
$\rho v_{\infty}^2$,
is transferred to the pressure of accelerated particles, $P_{cr}=\frac{4 \pi}{3} \int_{p_{\rm min}}^{p_{\rm max}} \text{d}p~p^3v(p)f(p)$, where $\rho$ is the density, $v(p)$ the particle velocity, $p_{\rm min}$ and $p_{\rm max}$ are the minimum and maximum momentum of accelerated particles. 
As long as  $p_{\rm min} \lesssim m_{\rm p} c$, its precise value is not relevant for determining the CR pressure.

 The ram pressure at the FS and TS, respectively expanding in the external medium and in the dense UFO wind read: 
 \begin{equation}
    P_{\rm ram}(t) = \left\{
    \begin{array}{ll}
        m_{\rm p}~n_0~v_{\rm FS}^2 & \mbox{[FS]} \\
        \frac{\dot{M}}{4\pi~R_{\rm TS}^2} v_{\infty}& \mbox{[TS]}
    \end{array}
\right.
\end{equation}
where $R_{\rm TS}$ is the termination shock radius. We assume that the UFO is sufficiently evolved such that the velocity of the TS satisfies $v_{\rm TS} \ll v_{\infty}$. The typical energy available at each shock can then be estimated by considering the volumes potentially filled with accelerated particles $4 \pi R_{\rm TS}^2 l_{\rm TS}$ and $4 \pi R_{\rm FS}^2 l_{\rm FS}$ associated to the TS and FS, respectively, where $l_{\rm TS} \sim R_{\rm FS} - R_{\rm TS}$ and $l_{\rm FS} \sim R_{\rm FS} - 0.9 R_{\rm FS}$.

 \begin{equation}
 \label{eq:E}
    E \sim \left\{
    \begin{array}{ll}
        10^{54} \left( \frac{n_0}{100 \text{cm}^{-3}}\right) \left(\frac{v_{\rm FS}}{0.1 \; c} \right)^2 \left( \frac{R_{\rm FS}}{3 \; \text{pc}}\right)^2 \left( \frac{l_{\rm FS}}{0.3 \text{pc}} \right) \text{erg} & \mbox{[FS]} \\
       10^{54} \left( \frac{\dot{M}}{1 M_{\odot}/\text{year}} \right)\left( \frac{v_{\infty}}{0.1 \; c} \right)  \left(\frac{l_{\rm TS}}{3 \text{pc}} \right)\text{erg}& \mbox{[TS]}
    \end{array}
\right.
\end{equation}
This illustrates that, although the TS is often regarded as the primary site of particle acceleration, the FS can also contribute significantly, especially at early time. It should also be emphasized that the estimates in Eq.~\eqref{eq:E} are time dependent, but the temporal evolution is mild; for instance, at the FS one finds $E \propto t^{-2/5}$. Moreover, the strength of the forward shock is governed by the average properties of the external medium, and efficient radiative losses in the shocked ambient gas may cause the shock to become radiative.
Although UFOs are observed to be transient and structured on short timescales near the central engine \citep{2025xrismflare}, such variability does not necessarily translate into rapidly evolving conditions at larger scales. The termination and forward shocks are governed by the time-averaged wind properties and evolve on dynamical timescales much longer than the events observed close to the central engine. We therefore adopt a steady state description based on ranges of UFO parameters to model the associated VHE emission.
In the following, we assume that particle acceleration occurs at a generic shock, without specifying whether it is the TS or the FS. The comparable energy budget of the two shocks (Equation~\eqref{eq:E}), supports such an assumption and makes our results on the hadronic byproduct general. We characterize the size of this shock by its radius, $R_{\rm sh}$, and consider that the accelerated particles occupy a volume corresponding to a fraction $\beta \simeq 0.5$ of the spherical volume $4\pi R_{\rm sh}^3/3$. This region is taken to be responsible for the production of the resulting gamma-ray and neutrino emission.

At this shock, the maximum energy of accelerated particles is tightly connected to the level of magnetic field around the shock. In the presence of efficient acceleration, the streaming of accelerated particles is likely to play a crucial role in the excitation of instabilities in the plasma, and to lead to a substantial amplification of the magnetic field especially at forward shocks. While at termination shocks a high level of turbulence and magnetic field amplification is naturally expected as a result of compression. 
Regardless of the process(es) at stake setting the average value of the magnetic field at the shock, we assume that a fraction $\xi_{\rm B}$ of the shock ram pressure is converted into magnetic energy density~\citep{1999Miniati_magnetic}: 
\begin{equation}
    U_B=\frac{B^2}{8\pi}=\xi _{\rm B} \rho v_{\infty}^2,
\end{equation}
with $\xi _{\rm B}\approx 10^{-4}$, typically resulting in magnetic field values on the order of hundreds of $\mu$G. 

We assume a magnetic conversion efficiency somewhat lower than~\citep{peretti2023diffusive}. This, besides resulting in a lower maximum energy for protons as discussed below, is for us a conservative assumption that allows the co-accelerated electrons not to cool too rapidly via synchrotron, thereby allowing for a potential contamination due to inverse Compton that we discuss in \S~\ref{Section: Results}.

The associated maximum energy attainable through DSA is set by the Hillas criterion~\citep{Hillas_origin_UHECR} :

\begin{equation}
     E_{\rm max} \approx 38 \, \left(\frac{r_{\rm sh}}{1~\rm pc}\right)^{-1} \, \left(\frac{v_{\infty}}{0.1~\rm c}\right) \, \left(\frac{\xi_{\rm B}}{10^{-4}}\right) \, \left(~\frac{\dot{M} }{0.1 ~\rm M_\odot/yr}\right) \, \text{PeV},
\label{eq:hillas}
\end{equation}
Our conservative choice of a low $\xi_{\rm B}$ should not be interpreted as excluding UFOs as potential sources of UHECRs - indeed $0.01<\xi_{\rm B}<0.1$ is compatible with $E_{\rm max} \gtrsim 10^{18} \, \rm eV$ - Rather it reflects our focus on a pessimistic scenario regarding potential electron contamination in the gamma-ray domain via inverse Compton emission. Such contamination would, in fact, be suppressed by stronger synchrotron losses for larger $\xi_{\rm B}$ values.

The main parameters of our model are summarized in Table~\ref{tab:parameters}. Other physical quantities such as the mass-loss rate, velocity of the outflows and distance of the UFO winds are taken from the literature~\citep{Ehlert_2024_UFO_UHECR}.

\begin{table}[]
\begin{tabular}{l|p{7.5cm}}
\hline
$\alpha$ &  slope of particles accelerated at the shock  \\
$\xi_{\rm CR}$ &  efficiency of acceleration at the shock \\
$\xi_{\rm B}$ & fraction of ram pressure transferred to magnetic field  \\
$n_0$ &  target material density \\
$R_{\rm sh}$ & typical size of the accelerating region\\
\hline
\end{tabular}
\caption{Parameters of the model}
\label{tab:parameters}
\end{table}

\begin{figure*}[!h]
    \centering
    \subfloat{
        \includegraphics[width=0.34\textwidth]{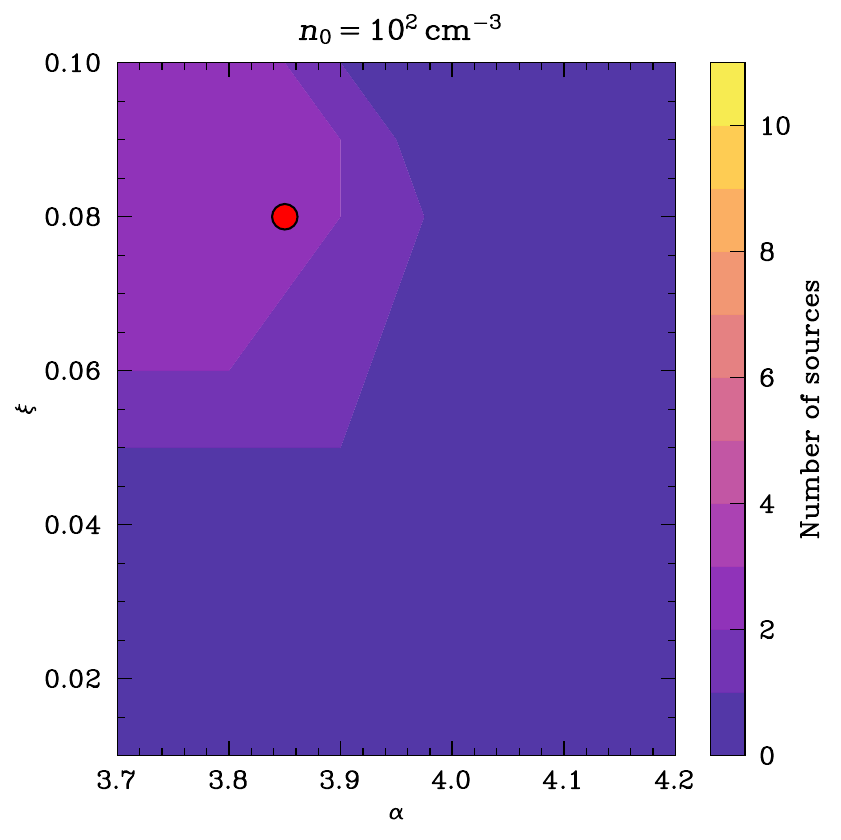}
        \label{fig:subfig1}
    }
    \subfloat{
        \includegraphics[width=0.53\textwidth]{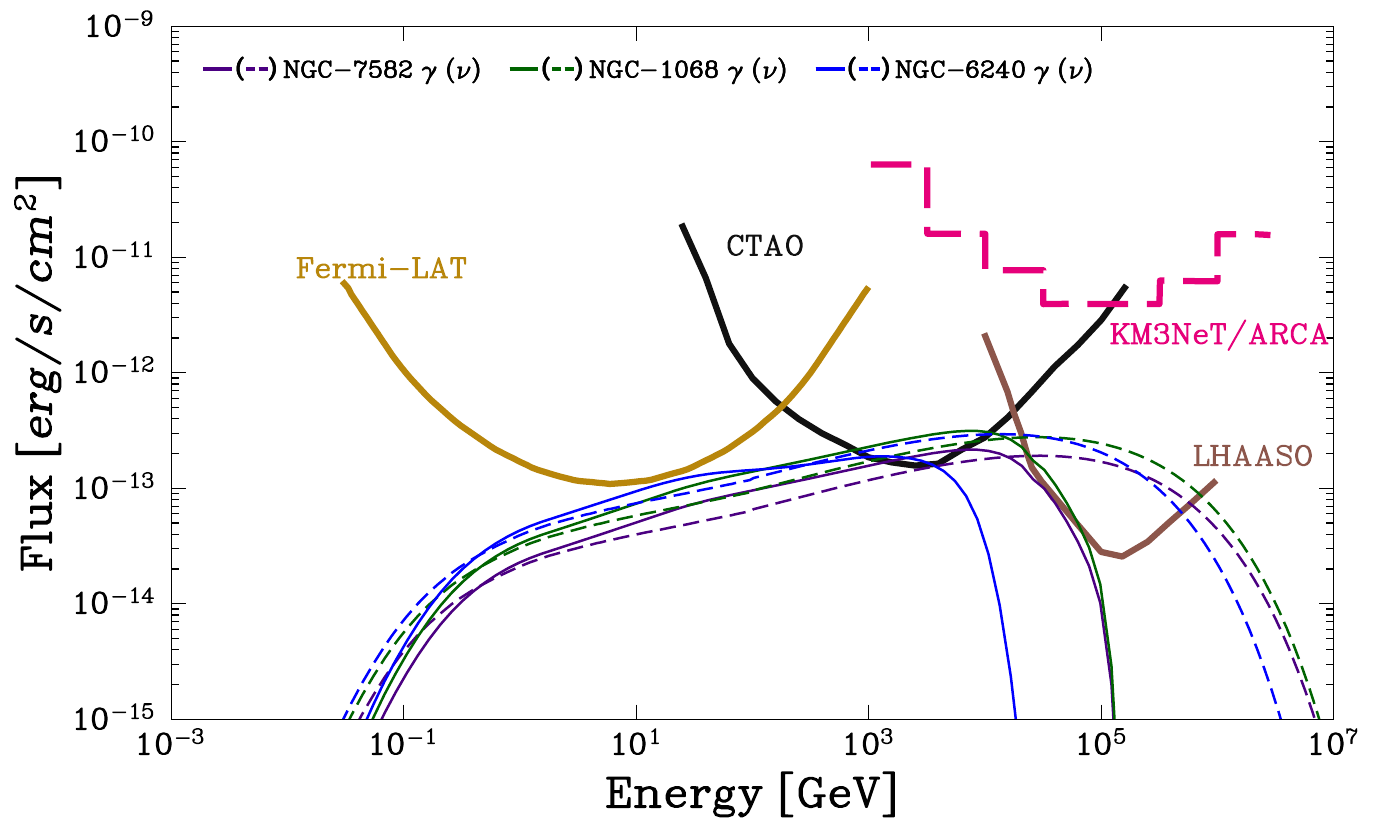}
        \label{fig:subfig2}
    }
    \vspace{0.5cm}
    \subfloat{
        \includegraphics[width=0.34\textwidth]{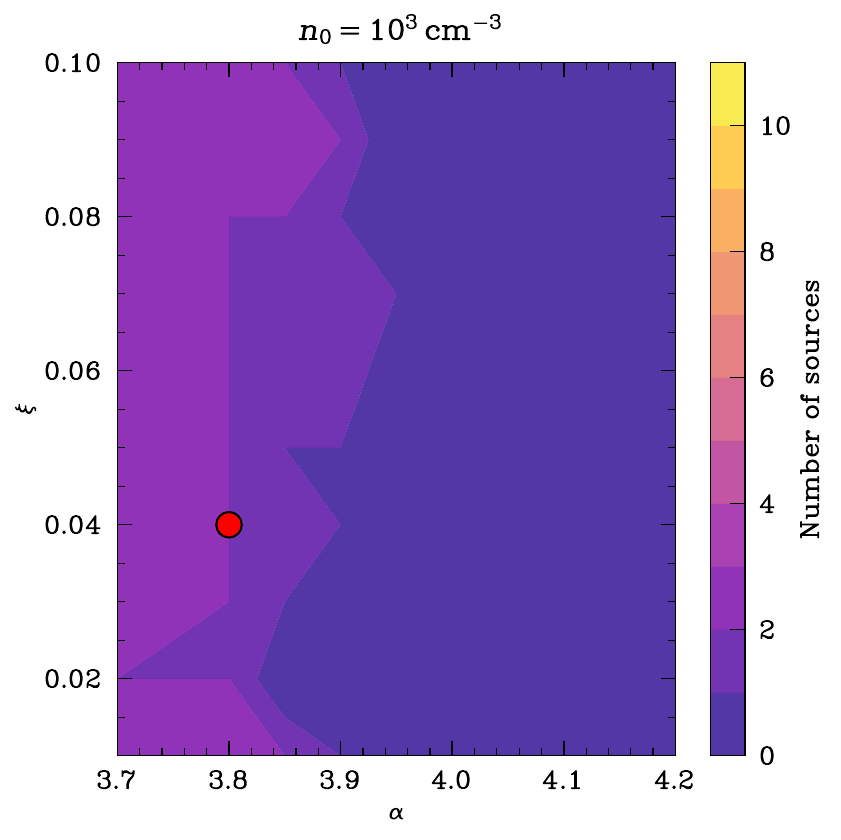}
        \label{fig:subfig3}
    }
    \subfloat{
        \includegraphics[width=0.53\textwidth]{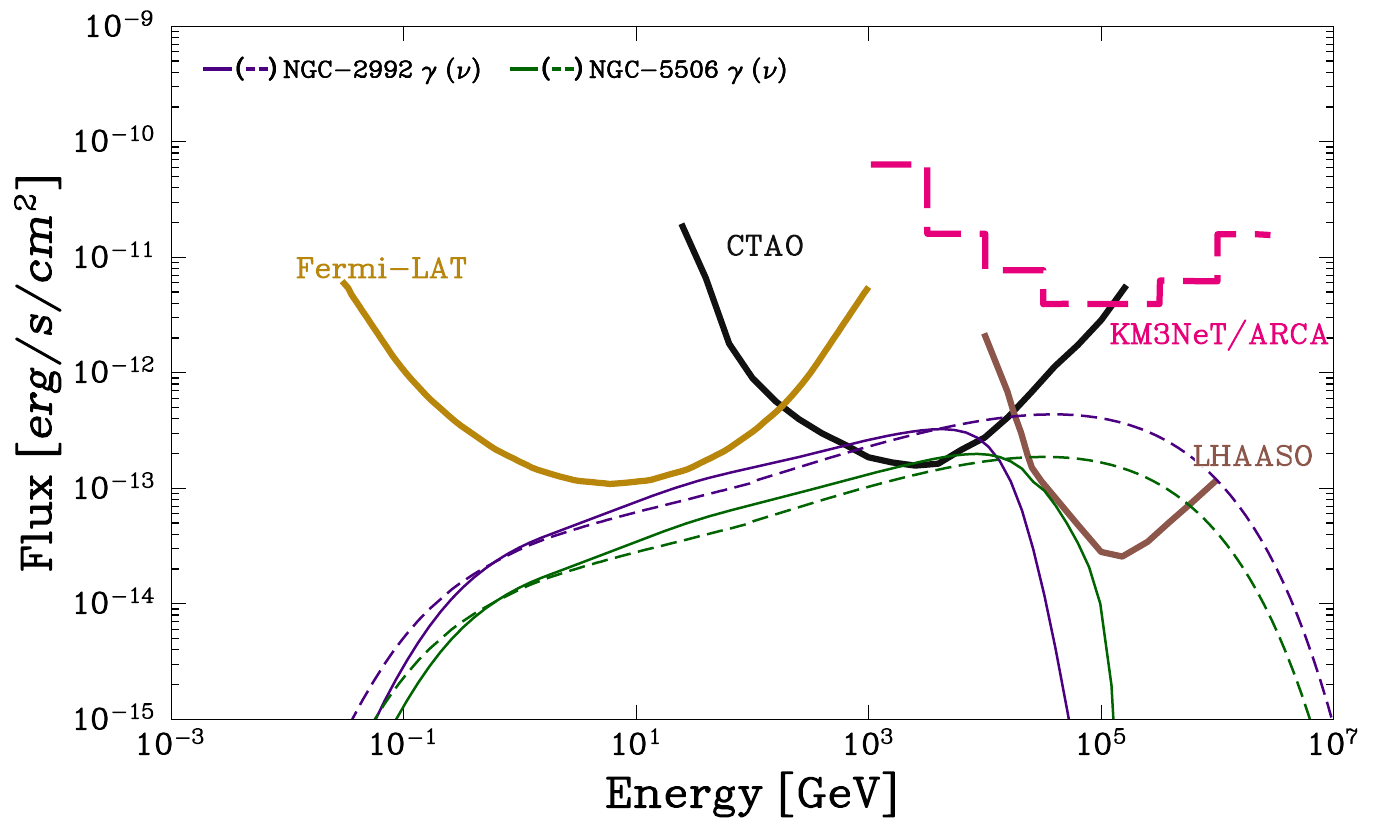}
        \label{fig:subfig4}
    }
  \vspace{0.5cm}
    \subfloat{
        \includegraphics[width=0.34\textwidth]{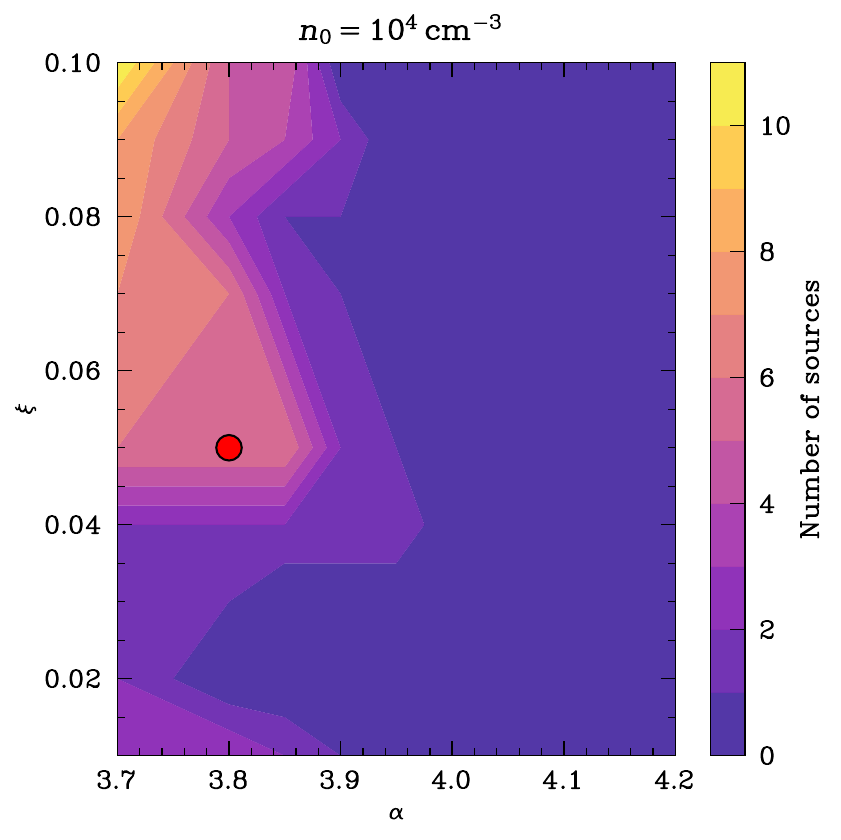}
        \label{fig:subfig3}
    }
    \subfloat{
        \includegraphics[width=0.53\textwidth]{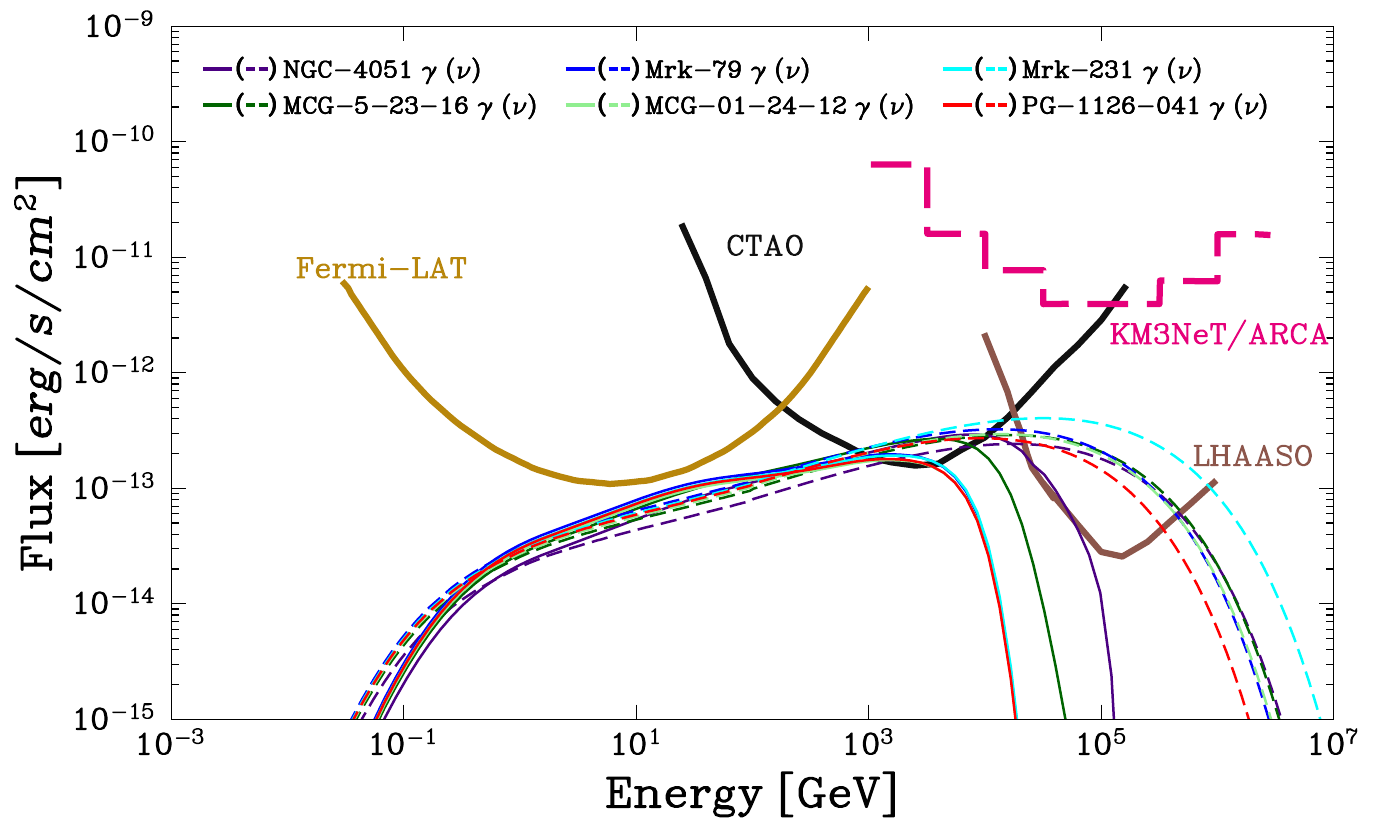}
        \label{fig:subfig4}
    }

    \caption{Number of UFOs with a gamma-ray signal above the typical sensitivity of the CTAO (50 hours) and below the Fermi-LAT sensitivity. The shock radius is $R_{\rm sh}= 1$ pc. Panels from top to bottom illustrate different assumptions of external density: $n_0 = 10^{2}-10^{3}-10^{4} \mathrm{cm} ^{-3}$. Sensitivities of Fermi-LAT (14 years), LHAASO (5 years) and KM3NeT/ARCA (10 years) are shown.}
    \label{fig:6plots}
\end{figure*}

\begin{figure*}[!h]
    \centering
    \subfloat{
        \includegraphics[width=0.34\textwidth]{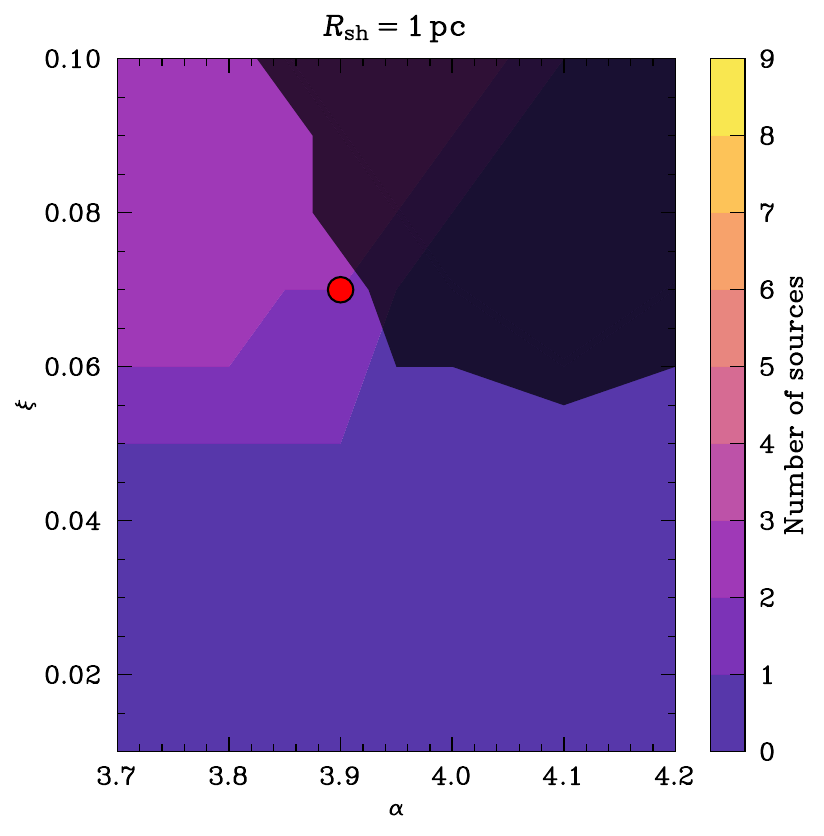}
        \label{fig:subfig1}
    }
    \subfloat{
        \includegraphics[width=0.53\textwidth]{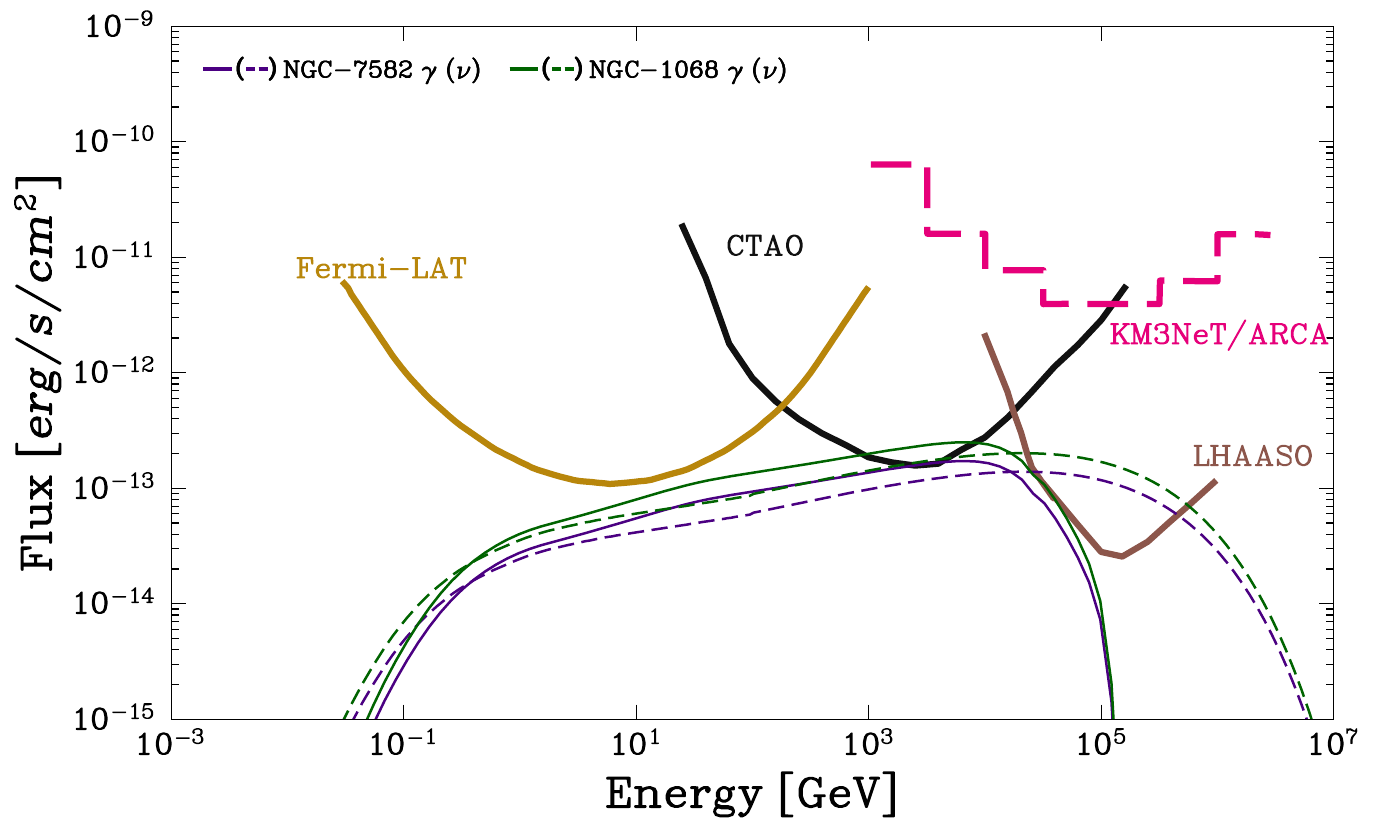}
        \label{fig:subfig2}
    }
  \vspace{0.5cm}
    \subfloat{
        \includegraphics[width=0.34\textwidth]{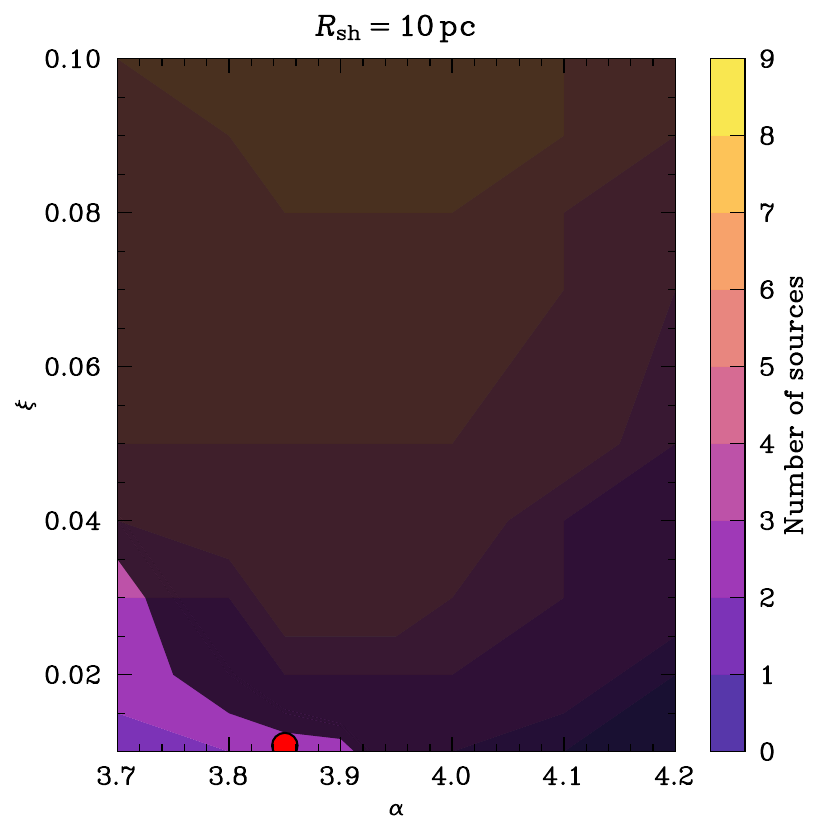}
        \label{fig:subfig3}
    }
    \subfloat{
        \includegraphics[width=0.53\textwidth]{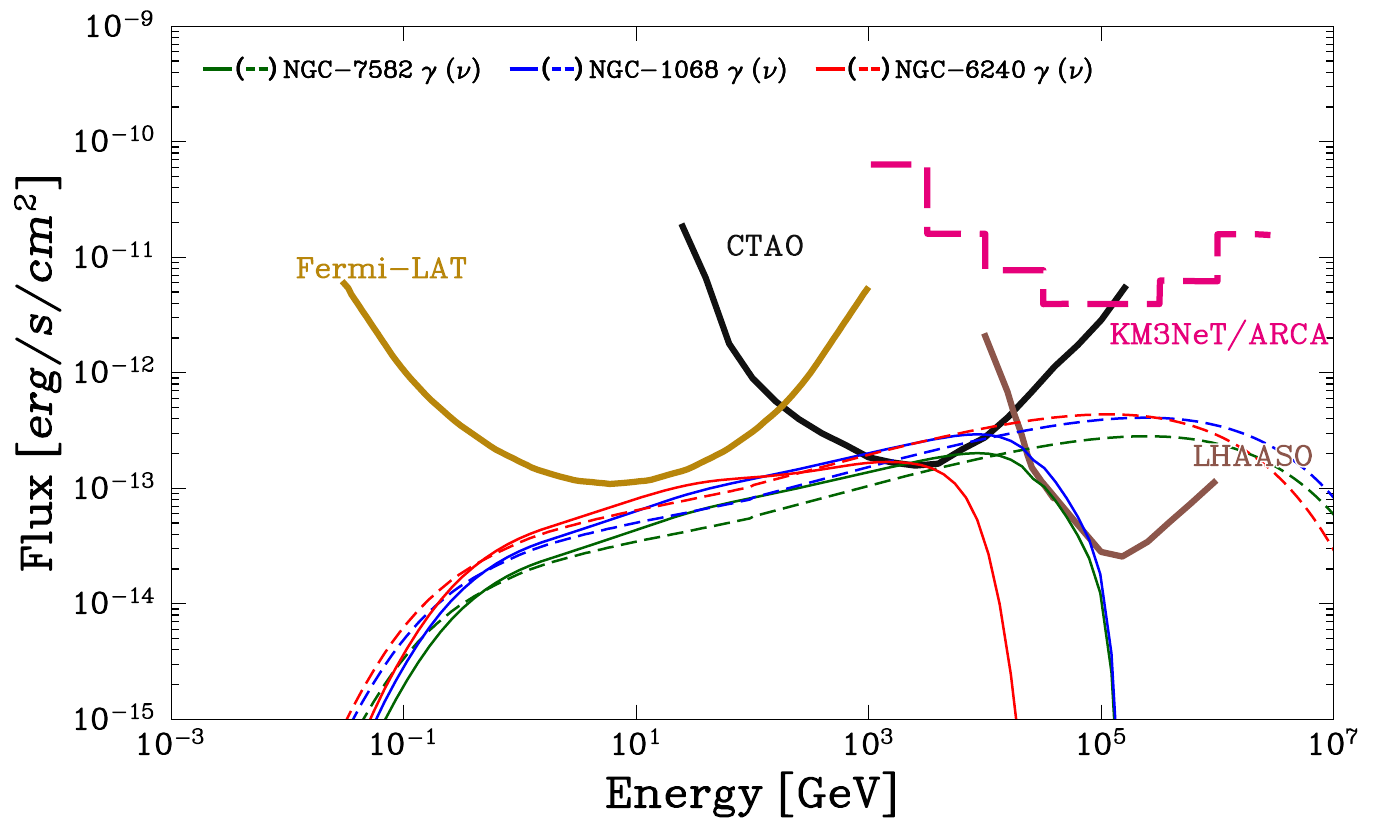}
        \label{fig:subfig4}
    }

    \caption{Number of UFOs with a gamma-ray signal above the typical sensitivity of the CTAO (50 hours) with dark areas for parameters that lead to a Fermi detection. The target density is $n_0 =10^{2} \mathrm{cm}^{-3}$. Sensitivities of Fermi-LAT (14 years), LHAASO (5 years) and KM3NeT/ARCA (10 years) are shown.}
    \label{fig:6 plots bis}
\end{figure*}

\subsection{Gamma rays and neutrinos}

The interactions of accelerated hadrons with protons and nuclei of the CNM result in the production of neutral and charged pions. 
{In turn}, gamma rays result from the decay of neutral pions, while neutrinos are the final products, together with electrons and positrons, of the decay of charged pions. 
We do not distinguish between neutrino flavors as the oscillation en route to Earth results approximately in flavor equipartition. The corresponding emissions are computed following~\citet{gamma_pp_Kelner} and~\citet{Kafexhiu_2014}.

The gamma-ray signal from pion decay is computed with \textit{Naima} code~\citep{Khangulyan_naima,gamma_pp_Kelner,Kafexhiu_2014}, assuming the proton distribution described by Eq.~\ref{protons distrib}. We neglect the leptonic contamination to the gamma-ray flux and we comment about the robustness of our assumption at the end of Section~\ref{Section: Results}.

The gamma-ray signal can be substantially degraded due to the interaction of gamma rays with low-energy photons through the two-photon process, creating electron-positron pairs. 
The absorption is taken into account considering two distinct phases of the photon propagation: 1) in the production region, high-energy gamma rays interact with the AGN background photon field; 2) outside the source, gamma rays interact with the photons of the cosmic microwave background (CMB) and of the extragalactic background light (EBL).
To account for 1), three AGN photon field components are considered, as a combination of different black bodies corresponding to two physical parts of the AGN: the dusty torus~\citep[$\approx10^3\,\mathrm{K}$][]{2011Mullaney}, the accretion disk ($\approx10^4\,\mathrm{K}$) and additionally a power law for the X-corona as in~\citet{2004Marconi}. The  absorption on CMB and EBL depends on the redshift of the source and we adopt the prescription described in~\cite{EBL_Franceschini_2017}.

\section{Results}
\label{Section: Results}

The main goal of our work is to discuss the prospects of detection of UFOs in the gamma-ray domain with present and next generation instruments such as the CTAO or LHAASO. 
We present the gamma rays and neutrinos expected for the list of 82 UFOs selected in~\citet{Ehlert_2024_UFO_UHECR}, following the methodology presented in the previous section.

Setting aside the previously discussed cases of NGC 1068 and NGC 4151, the lack of Fermi–LAT detections of UFOs is not straightforward to interpret. This could either indicate that the physical conditions in each source hinder GeV emission, or that, intrinsically, UFOs as a population are inefficient producers of GeV gamma rays.
As the non-detection remains difficult to interpret, we adopt two approaches. In the first, we examine all sources individually and investigate under which conditions they can be detected by CTAO while remaining undetected by Fermi. In the second, we study the sample as a whole and explore the regions of parameter space that could lead to a CTAO detection, while remaining consistent with the non-detection of all UFOs by Fermi.

In our first approach, we perform an exploration of the parameter space ($\alpha$, $\xi_{\rm CR}$, $n_{\rm 0}$, $R_{\rm sh}$ and $B$) and compute the gamma-ray and neutrino signal expected from each source. 
We focus on sources whose gamma-ray spectra exceed the sensitivity of the CTAO, and remain below the sensitivity of Fermi-LAT to account for the fact that no clear UFO detection has been claimed by Fermi-LAT besides potentially NGC~4151 and NGC~1068. 

The number of expected UFOs for typical values of $\alpha$, $\xi_{\rm CR}$ and $n_0$ are illustrated in Fig.~\ref{fig:6plots}. 
We show (from top to bottom) three different assumptions for the external target density. On the left hand side of each panel we display the parameter space of the detectable sources.
The expected gamma-ray and neutrino fluxes for the sources of interest are illustrated on the right sides. 
The typical differential sensitivities of Fermi-LAT, the CTAO and the LHAASO are also displayed for comparison.
In our sample of best UFOs candidates, an index $\alpha \lesssim 3.9$ and an efficiency $\xi_{\rm CR}\gtrsim 0.05$ lead to at least one potential detection in the very-high-energy domain, regardless of the value assumed for the target density $n_0$. 
As expected, the number of expected sources increases with $n_0$, $\xi_{\rm CR}$, $R_{\rm sh}$ (increasing the CR density and target density) and with harder spectra (smaller values of $\alpha$).

In our second approach, we treat the sample of UFOs as a whole and investigate the regions of parameter space that could lead to a detection with CTAO, while remaining consistent with the fact that all UFOs are undetected with Fermi-LAT. Results are illustrated in Figure~\ref{fig:6 plots bis}. 
The difference between the two approaches can be seen by comparing the top panels of Fig.~\ref{fig:6plots} and top panels Fig.~\ref{fig:6 plots bis}, which correspond to a size of the accelerating zone $R_{\rm sh}=1~\mathrm{pc}$ and density $n_0 = 10^2$ cm$^{-3}$. 

Our second approach relies on a more drastic criterion, thus significantly reducing the number of detectable UFOs, and the interesting regions of the parameter space for a detection in the VHE range. As shown in Fig.~\ref{fig:6 plots bis}. Even with this stricter criterion, several UFOs remain detectable across specific regions of the  parameter space explored for both $R_{\rm sh}=1~\mathrm{pc}$ and $R_{\rm sh}=10~\mathrm{pc}$.

Both approaches suggest that at least one UFO with a hadronic TeV component could be observable, while still being consistent with the lack of UFO detections in the GeV range.
The best UFOs candidates for a detection in the TeV are reported in Table~\ref{tab:names_fig2}.
The most promising sources are the nearest ones, for which the gamma-gamma absorption on the CMB and EBL is limited. Among these, those with outflow speeds $v_{\rm outflow}\gtrsim  0.1~\rm c$ are particularly favorable. 

It is worth noting that NGC~4151, a Seyfert galaxy with evidence of hosting an Ultra-Fast Outflow, does not appear among our TeV-detectable candidates. This results from our selection criterion that excludes all sources above the Fermi–LAT sensitivity. However, its potential detection by Fermi–LAT, consistent with the properties of the UFO~\citep{ngc4151} suggests that it could be a relevant target for further observations in the TeV domain. 
The case of NGC 1068 also warrants special attention. Although it is detected by Fermi–LAT and would thus lie above our selection threshold, the origin of its GeV emission remains ambiguous. The emission is likely a composite of AGN and circumnuclear components in a highly opaque environment, with indications that a putative UFO could contribute to, and potentially contaminate, the observed flux \citep{peretti2023diffusive}. In this work, we explore scenarios in which NGC 1068 would remain undetected by Fermi–LAT but detectable with the CTAO. Such cases imply that the GeV gamma rays observed from the host would not originate from the UFO shocks.

Having defined our sample, we now comment on the modeling of the expected emission mechanisms. In particular, we evaluate the possible leptonic contamination to the VHE spectra.

The bulk of the UFO kinetic power is expected to be thermalized at the shock. 
For a strong, adiabatic, non-relativistic shock ($\gamma=5/3$), the immediate downstream temperature is given by the Rankine-Hugoniot jump conditions:
\begin{equation}
T_2 = \frac{3}{16}\frac{\mu m_p}{k_B} v_s^2 
\approx 1.2 \times 10^{10}
\left(\frac{\mu}{0.6}\right)
\left(\frac{v_s}{0.1c}\right)^2 \ \mathrm{K},
\end{equation}
corresponding to $ 
k_B T_2 \approx 1.1
\left(\frac{\mu}{0.6}\right)
\left(\frac{v_s}{0.1c}\right)^2 \ \mathrm{MeV}.
$
In a collisionless shock, this temperature primarily reflects the ion temperature immediately behind the shock, since ions receive most of the dissipated kinetic energy. 
The electron temperature can be parameterized as: 
$T_e = \beta_{ep} T_p$
where $\beta_{ep} \equiv T_e/T_p \ll 1$ for high Mach-number shocks. 
Observations of collisionless strong shocks and kinetic simulations suggest $\beta_{ep} \sim 10^{-2}-10^{-1}$, with a conservative lower bound of order $m_e/m_p$ \citep[e.g.,][]{ghavamian2007,ghavamian2013,vink2015}

Therefore although shock heating converts a substantial fraction of the kinetic power into thermal energy, the resulting thermal X-ray emission (bremsstrahlung and, at lower $T_e$, line emission) is expected to remain subdominant relative to the intrinsic AGN X-ray luminosity. The compact AGN corona typically dominates the hard X-ray output, while the shocked region has a limited emission measure due to its thin shell geometry. In addition, circumnuclear absorption and geometrical effects can attenuate extended thermal emission. Therefore, significant shock heating does not necessarily imply a dominant or readily identifiable thermal X-ray component from the UFO shock itself.

\begin{figure}[h!]
   \centering
   \includegraphics[scale=0.55]{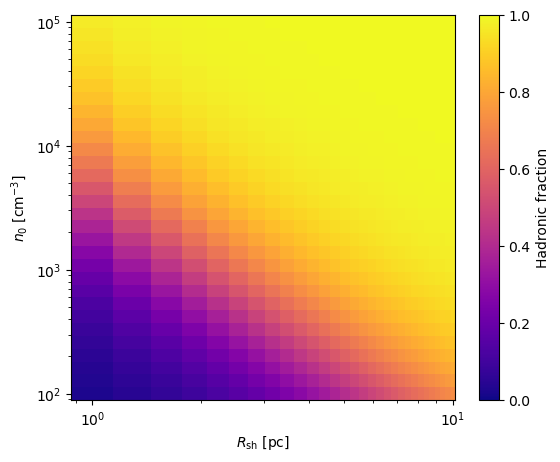}
      \caption{Ratio of hadronic gamma rays over hadronic + leptonic gamma rays $R_{\rm hadronic}= \frac{F_{\rm had}( > 100 ~\text{GeV})}{F_{\rm had}( > 100 ~\text{GeV})+F_{\rm lep}( > 100 ~\text{GeV})}$ for various values of the shock radius $R_{\rm sh}$ and ambient density $n_0$. In this figure, the bolometric luminosity is fixed at $L_{\rm bol} = 10^{43}~\rm erg/s$, and the electron-to-proton ratio is set to  $K_{\rm ep} = 10^{-3}$. The leptonic gamma-ray flux is computed for Inverse Compton with \textit{Naima} code~\citep{Kafexhiu_2014,2010_Aharonian} assuming that the electron population follows the same energy distribution as the protons, with normalization $K_{\rm ep}$ at 1 GeV. The AGN photon field considered is described in Sec.~\ref{Section: Model}, and the maximum energy of electrons is estimated by equating the acceleration timescale to the smaller of the losses timescale and the UFO age.}
         \label{fig:Kep}
   \end{figure}

In fact, accelerated electrons can also contribute to gamma-ray emission through inverse Compton scattering, synchrotron radiation, and Bremsstrahlung.
However, this contribution depends on several physical quantities, such as the magnetic field strength, the ambient photon density, and the amount of accelerated electrons. 
To estimate the leptonic gamma-ray contribution, we consider an electron population following the same distribution as the protons, with 
$A_{\rm e} = K_{\rm ep} A$, 
where $A$ is the proton normalization introduced in Eq.~\ref{protons distrib}, and $K_{\rm ep}$ is the electron-to-proton ratio. 
The value of $K_{\rm ep}$ is highly uncertain and is typically assumed to lie in the range $10^{-5}$-$10^{-2}$~\citep{morlino_Tycho,Park_protons_electrons,Batzofin_SNR}. 
As shown in Fig.~\ref{fig:Kep}, provided that $K_{\rm ep} < 10^{-3}$, the leptonic gamma-ray emission remains negligible compared to the hadronic signal.

For photon energies in the TeV range, Compton scattering occurs deep in the Klein-Nishina regime, where the cross section is strongly suppressed relative to the Thomson value ($\sigma_{\rm KN} \ll \sigma_T$). 
Taking a representative density $n_e \sim 10^3 \ \mathrm{cm^{-3}}$ and a characteristic size $R \sim 1 \ \mathrm{pc} \simeq 3 \times 10^{18}$ cm, the Thomson optical depth would be $
\tau_T \sim n_e \sigma_T R \sim 2 \times 10^{-3}.$
In the Klein-Nishina regime, the effective cross section is further reduced by orders of magnitude, leading to: 
$\tau_{\rm KN} \lesssim 10^{-5}-  10^{-4}$. 
For the densities ($n_0 \lesssim 10^2-10^4$ cm$^{-3}$) and the lenght scales (pc) considered in this work the corresponding optical depth remains negligible ($\tau_{\rm KN} \ll 1$) . 
Thus, significant attenuation or energy degradation of VHE photons via Compton scattering would require electron column densities several orders of magnitude larger than expected in UFO environments.

For the most luminous AGN, sustaining hadronic dominance at the larger characteristic radii requires higher ambient gas densities, because the inverse Compton luminosity increases with both the electron normalization and the external photon energy density ($\propto K_{\rm ep}\,u_{\rm ph}$), whereas the hadronic component scales with the target gas density ($\propto n_{0}$). Conversely, for all lower-luminosity AGN considered here, hadronic dominance remains a robust expectation across the full range of ambient densities and shock radii explored, $(n_{0},\,R_{\rm sh})$.
Consequently, in the scenarios considered here, we neglect the contribution of electrons to the total gamma-ray flux.

\begin{table}[h!]
\centering
\caption{Best UFO candidates associated  and their properties to Fig.~\ref{fig:6plots}. }
{\small

\begin{tabular}{lllll}
\hline
Sources & Distance [Mpc] & $v_{\rm outflow}$ [$c$] & Type \\
\hline
NGC~4051              &   8 & 0.128 & Seyfert 1.5  \\
MCG‑5‑23‑16$^{*}$     &  38 & 0.116 & Seyfert 1.9  \\
\textbf{NGC~7582}$^{*}$        &  21 & 0.285 & Seyfert 2  \\
Mrk~79                &  94 & 0.091 & Seyfert 1  \\
3C~111                & 209 & 0.083 & FRII      \\
IC~5063$^{*}$         &  47 & 0.311 & Seyfert 2  \\
IRAS~05054+1718(W)    &  77 & 0.176 & Seyfert 1 \\
MCG‑01‑24‑12$^{*}$    &  85 & 0.098 & Seyfert 2  \\
MCG‑03‑58‑007$^{*}$   & 137 & 0.193 & Seyfert 2  \\
Mrk~231               & 179 & 0.241 & Seyfert 1      \\
Mrk~273               & 162 & 0.265 & Seyfert 2 \\
Mrk~279               & 128 & 0.220 & Seyfert 1  \\
\textbf{NGC‑1068}$^{*}$        &  17 & 0.277 & Seyfert 2  \\
NGC~2992              &  34 & 0.298 & Seyfert 1  \\
NGC~5506              &  25 & 0.247 & Seyfert 1  \\
\textbf{NGC~6240}              & 107 & 0.125 & Seyfert 2 \\
PG~0844+349           & 274 & 0.211 & Seyfert 1       \\
PG~1126‑041$^{*}$     & 256 & 0.064 & NLS 1      \\
SWIFT~J2127.4+5654    &  64 & 0.231 & NLS 1  \\
\hline
\end{tabular}%

}

\tablefoot{In bold are the best candidates found in the second approach illustrated in Fig.~\ref{fig:6 plots bis}. Southern‐hemisphere sources are marked with $^{*}$.}
\label{tab:names_fig2}
\end{table}

\section{Conclusions}
\label{Section: Conclusions}

For a list of nearby selected UFOs~\citep{Ehlert_2024_UFO_UHECR}, we model particle acceleration at the strong shocks produced by these outflows, and estimate the subsequent expected gamma-ray and neutrino signals.
Our main conclusions can be listed as follows. 

\begin{enumerate}
    \item Several UFOs are expected to be detectable by next-generation TeV gamma-ray and neutrino observatories. The number of detections depends on the physical parameters that govern particle acceleration at the UFO shocks. By exploring the parameter space, we find that harder spectra ($\alpha \lesssim 3.9$) lead to an increased number of UFOs detectable in the TeV range, while remaining undetected in the GeV range. The size of the acceleration region ($R_{\rm sh}$), the efficiency of particle acceleration ($\xi_{\rm CR}$), and the target density ($n_{0}$) all directly increase the expected number of detections. 

     \item In our first approach, we consider all UFOs with gamma-ray luminosities above the CTAO sensitivity in the TeV range, and exclude those lying above the Fermi-LAT sensitivity. In this framework, we find that $\alpha \lesssim 3.9$ is required to ensure a TeV detection. For lower target densities ($n_{0} = 10^{2}$ cm$^{-3}$), we further find that a cosmic-ray acceleration efficiency of $\xi_{\rm CR} \gtrsim 0.05$ is necessary.

    \item In our second approach, we exclude the regions of parameter space that would lead to results in tension with the non-detection of UFOs by Fermi-LAT. Under these conditions, we find that a few UFOs could be detectable with the CTAO. The spectra of the detectable sources are found to be hard ($\alpha \leq 4.0$) for densities $n_{0} \geq 10^{2}$ cm$^{-3}$. This approach, being more stringent than the first one, leads to a reduced number of detections. Higher densities ($n_{0} \gtrsim 10^{3}$ cm$^{-3}$) rapidly shrink the allowed regions of parameter space, ultimately resulting in no detectable sources.

    \item We find that spectra harder than $p^{-4}$ are required for detection in the VHE domain. Such spectra can, a priori, arise from several effects, including non-linear processes due to efficient particle acceleration at the shock, or the escape of particles from the acceleration region. Radiative cooling behind the shocks, which can increase the compression factor, could also lead to harder spectra~\citep{2025_Cristofari_Radiative_shocks}. 
    The presence of clumps, that has been suggested by recent observations~\citep{2025_xrism_clumps}, could also play a role in hardening the gamma-ray spectrum~\citep{2014gabici_aharonian}.
    A detection in the VHE domain could therefore indicate that one or more of these effects are at play.

    \item Should any UFOs be detected in the VHE gamma-ray domain with hard spectra, this would indicate that additional processes not accounted for in our study are at play — for instance, the contribution of accelerated electrons. Although their role in gamma-ray production is expected to be subdominant given the high densities and magnetic fields involved, it will be investigated in a forthcoming study.
    
    \item In both approaches, we investigate the prospects for detection with next-generation neutrino observatories, and find that the prospects for detection with KM3NeT/ARCA in  10 years are bleak.
    
    \item The best candidates for detection in the VHE domain naturally include some of the closest UFOs in the selected sample: NGC~7582, NGC~4051,  NGC~5506, NGC~2992 and NGC~6240. All of them located at $\lesssim 270~$Mpc, and with outflow velocities $\gtrsim 0.05~\rm c$. 

     \item A non detection of UFO shocks would imply either that shocks do not form or that they are radiatively inefficient. The former would challenge current expectations for magnetohydrodynamic outflow propagation, while the latter would instead constrain the efficiency of dissipation and particle acceleration at such shocks. In this sense, non-detections primarily inform the microphysics of interactions in the UFO environment rather than implying the absence of shocks.

    \item Comparing the predictions of our model with future gamma-ray observations of the UFO population will offer a novel approach to probing particle acceleration in these outflows. Furthermore, the global properties of the UFO population-such as source count, luminosity distribution, distances, and spectral indices - can help place tighter constraints on particle acceleration mechanisms at strong, sub-relativistic collisionless shocks.

\end{enumerate}

In a follow-up investigation we plan to explore the multiwavelength impact of leptons in UFOs as well as a dedicated study of potential and nearby UFO hosts detected in gamma-ray by Fermi-LAT such as NGC~1068 and NGC~4151. It is important to note that NGC~1068 has been proposed as a potential UFO host~\citep{2024hypothesis_NGC1068}, but this interpretation remains under debate and lacks definitive evidence.

\begin{acknowledgements}
      This work was supported by the PSL Starting Grant GALAPAGOS. 
\end{acknowledgements}

   \bibliographystyle{aa} 
   \bibliography{biblio.bib} 

\end{document}